\documentstyle[12pt]{article}

\global\arraycolsep=1pt
\newcommand{\resection}[1]{\setcounter{equation}{0}\section{#1}}

\newcommand{\bel}[1]{\begin{equation}\label{#1}}                     
\newcommand{\bal}[1]{\begin{eqnarray}\label{#1}}                     
\newcommand{\Eq}[1]{(\ref{#1})}                                    
\newcommand{\be}{\begin{equation}}
\newcommand{\ee}{\end{equation}}
\newcommand{\ba}{\begin{eqnarray}}
\newcommand{\ea}{\end{eqnarray}}
\newcommand{\nn}{\nonumber \\}
\newcommand{\s}{\sigma}
\newcommand{\vac}{{\rm vac}}

\begin{document}

\begin{titlepage}

\begin{flushleft}
{\baselineskip=14pt 
\rightline{
 \vbox{\hbox{ITP-SB-98-14}
       \hbox{YITP-98-11}
       \hbox{February 1998}       }}}
\end{flushleft}

\vskip 2.0 cm

\begin{center}

{\large\bf The Determinant Representation 
            for a Correlation Function 
            in Scaling Lee-Yang Model} \\
\vskip 3.0cm

{ Vladimir E. Korepin$^{1}$
\footnote{e-mail: {\tt korepin@insti.physics.sunysb.edu}}
and 
Takeshi Oota$^2$}
\footnote{e-mail: {\tt toota@yukawa.kyoto-u.ac.jp}}\\[2em]

$^1${\sl Institute for Theoretical Physics, State University of New
           York at Stony Brook,\\
           Stony Brook, NY 11794--3840, U. S. A.}\\
$^2${\sl Yukawa Institute for Theoretical Physics, Kyoto
           University, Kyoto 606-01, Japan}\\

\end{center}

\vskip24pt

\begin{abstract}
\noindent
 We consider the scaling Lee-Yang model. It corresponds to the unique 
perturbation of the minimal CFT model $M_{2,5}$. This is not a unitary model.
We used known expression for form factors in order to  obtain a closed
expression for a correlation function of a trace of  energy-momentum
tensor. This expression is a determinant of an integral operator. 
Similar determinant representation were proven to be useful not only
for quantum correlation functions  but also in matrix models.
\end{abstract}

\end{titlepage}

\resection{Introduction}

The theory of massive, relativistic, integrable models is an important
part of modern quantum field theory \cite{ZZ}--\cite{N}. Scattering matrices
in these models  factorize into a product of two-body 
S-matrices \cite{ZZ}. Form factors can be calculated on the basis of
a bootstrap approach \cite{ZZ}--\cite{N}. 

The purpose of this paper is to
calculate correlation functions. As usual correlation function can be
represented as an infinite series of form factors contributions. 
In this paper we sum up all these contributions and obtain a closed 
expression for the correlation function of the energy-momentum tensor.
We follow an approach of the paper \cite{KS}. We 
introduce an auxiliary Fock space and auxiliary Bose fields (we shall 
call them dual fields).  These fields help us to represent the form 
factor decomposition of a correlation function in a form similar to 
``free fermionic'' case. This approach was developed in
\cite{K,Sl,K.B.I.}. Finally a correlation function is
represented as a vacuum mean value (in the auxiliary Fock space) of a
determinant of an integral operator \Eq{Ldet}.
Determinant representation  were proven to be useful not only
for quantum correlation functions \cite{BMW} - \cite{LLSS} but also 
in matrix models \cite{D} - \cite{FO}.
It helps for asymptotical analysis of quantum correlation functions.
Among other things this approach helped to calculate asymptotic of time
and temperature dependent correlation function in Nonlinear Schr\"odinger
equation \cite{KS1}. 

The scaling Lee-Yang model can be described by the 
$\phi_{1,3}$-perturbation of the non-unitary minimal model $M_{2,5}$ 
\cite{CM2}. The theory is known to be integrable and is described by 
factorizable scattering theory of only one kind of particle with mass $m$. 
The two-body scattering amplitude is given by \cite{CM2}
\be
S(\beta) = \frac{\sinh \beta + i\sin (2\pi/3)}{\sinh \beta - i\sin (2\pi/3)}.
\ee
The pole at $\beta = 2\pi i/3$ corresponds to a bound state pole. 
The residue of the pole is negative. So three point coupling is imaginary.
These facts come from non-unitarity of the theory. This S-matrix can be
obtained from breather-breather S-matrix of Sine Gordon \cite{S}.
The breather-breather S-matrix of Sine Gordon was first calculated in 
\cite{AK}.

Form factors for the trace of the energy-momentum tensor 
$\Theta = T^{\mu}_{\mu}/4$ is defined by the matrix element 
between the vacuum state $\langle \vac |$ and $n$ particle states
characterized by their rapidities $\beta_i$ ($i = 1, \ldots, n$):
\be
F_n ( \beta_1, \ldots, \beta_n ) =
\langle \vac | \Theta(0) | \beta_1, \ldots, \beta_n \rangle.
\ee
The multi-particle form factors $F_n$ was determined in \cite{S,Z}:
\bel{Iff}
F_n ( \beta_1, \ldots, \beta_n ) =
H_n Q_n(x_1, \ldots, x_n) 
\prod_{i<j} 
  \frac{ f( \beta_i - \beta_j ) }{ x_i + x_j },
\ee
where $x_i = e^{\beta_i}$, $i=1, \ldots, n$,
\bel{Hn}
H_n = - \frac{ \pi m^2 }{ 4 \sqrt{3} }
\left(
  \frac{ i 3^{1/4} }{ 2^{1/2} v(0) }
\right)^n.
\ee
The function $f(\beta)$ is given by
\be
f(\beta) = \frac{\cosh \beta - 1}{\cosh \beta + 1/2}
v(i\pi - \beta) v(-i \pi + \beta),
\ee
where
\be
v(\beta)=\exp\left(2 \int_0^{\infty}
\frac{dt}{t}e^{i \beta t/\pi} 
\frac{\sinh(t/2) \sinh(t/3) \sinh(t/6)}{\sinh^2 t} \right).
\ee
The function $f(\beta)$ has a single pole in the strip 
$0\leq {\rm Im}\beta <\pi$ at $\beta = 2\pi i/3$ and a single zero at
$\beta = 0$.

We shall use the  elementary symmetric polynomials $\s_k (x_1, \ldots, x_n)$,
\[
\prod_{j=1}^n (x+ x_j) = 
\sum_{k \in {\bf Z}} x^{n-k} \s_k(x_1, \ldots, x_n),
\]
It is important to remember that $\sigma_k$ is equal to zero if $ k<0 $
 or $ k>n $ .
The symmetric polynomials $Q_n(x_1, \ldots, x_n)$ arising in the formula
for form factors can be written as:
\bal{IQn}
Q_0 &=& 1, \nn
Q_1 &=& 1, \nn
Q_2 &=& \s_1, \nn
Q_n &=& \s_1 \s_{n-1} P_n, \ \ \ \ \  n \geq 3,
\ea
where
\be
P_n ={\det}_{n-3} \left( \Sigma_{ij} \right).
\ee
Here $\Sigma_{ij}$ is an $(n-3)$ by $(n-3)$ matrix with the entries
\bel{Sigij}
\Sigma_{ij} = \s_{ 3i - 2j + 1 }, \ \ \ \ 1 \leq i, j \leq n-3.
\ee
The index $n-3$ in the expression ${\det}_{n-3}$ denotes the dimension 
of the matrix $\Sigma_{ij}$.

After the Wick rotation to the Euclidean space, 
a correlation function of the operator $\Theta$ can be presented as an 
infinite series of form factors contributions
\bal{Icorff}
\langle \Theta(x) \Theta(0) \rangle &=&
\sum_{n=0}^{\infty} \int \frac{d^n\beta}{n!(2\pi)^n}
\langle \vac| \Theta(0) |\beta_1, \ldots, \beta_n \rangle
\langle \beta_n, \ldots, \beta_1 | \Theta(0) |\vac \rangle \nn
& &\times \exp\left[-m r \sum_{j=1}^n \cosh \beta_j \right],
\ea
where $r = (x^{\mu} x_{\mu})^{1/2}$.

In the present paper we sum up this series explicitly. 
Now let us present a  plan of the paper. 
Section 2 is devoted to a transformation of the determinants to a
form, which is convenient for summation.  
In section 3 we introduce auxiliary quantum operators---dual fields---in
order to factorize an expression for a correlation function and to represent
it in a form similar to ``free fermionic case". 
In section 4 we sum up the series \Eq{Icorff} into a Fredholm
determinant. 
In section 5 we use the Fredholm determinant representation 
for derivation of an asymptotic behavior of the  correlation function.

\resection{A transformation of the form factor}

A  determinant of a linear integral operator $I+V$ can be written
as
\bel{TdefFrdet}
\det(I+V)=\sum_{n=0}^\infty\int\,\frac{dx_1\cdots dx_n}{n!}
{\det}_n\left(
\begin{array}{ccc}
V(x_1,x_1)&\cdots&V(x_1,x_n)\\
V(x_2,x_1)&\cdots&V(x_2,x_n)\\
\cdot&\cdot&\cdot\\
V(x_n,x_1)&\cdots&V(x_n,x_n)
\end{array}\right).
\ee
Thus, in order to obtain a determinant representation for correlation
functions one need to represent the form factor expansion
\Eq{Icorff} in the form \Eq{TdefFrdet}. Determinants of integral operators,
which we consider also can be called Fredholm determinants.

The form factor \Eq{Iff} is  proportional to the  polynomials $Q_n$
\Eq{IQn}. The polynomial $Q_n$ for $n\geq 3$ is proportional to the
determinant of the matrix $\Sigma_{ij}$ \Eq{Sigij}. 

Let $M$ be the following $n$ by $n$ matrix:
\bel{Tmjk}
M_{ij} = \s_{ 3i - 2j - 1 }, \ \ \ \ 1 \leq i, j \leq n.
\ee
Note that $M_{1j} = \delta_{1,j}$, $M_{2j} = \delta_{1,j} \s_3 +
\delta_{2,j} \s_1$ and $M_{nj} = \delta_{n,j} \s_{n-1}$ for $j = 1,
\ldots, n$. Using the relation $M_{ij} = \Sigma_{i-2, j-2}$ ($i, j = 3, 
\ldots ,n-1$), it is possible to show that
\be
Q_n = {\det}_n M, \qquad n\geq 0.
\ee
The matrix $M_{ij}$ consists of $n^2$ different functions,
depending on the same set of arguments $x_1,\dots,x_n$.

The main goal of this and  next section is to transform the matrix
\Eq{Tmjk} to  such a form, that entries of a new matrix would be 
parameterized by a single function, depending on different sets of 
variables, (like $V(x_i,x_j)$ in \Eq{TdefFrdet})
\bel{Tidea}
M_{ij}\to \hat{D}_{ij}, \qquad \hat{D}_{ij} = \hat{D}(x_i, x_j).
\ee

In order to study correlation functions, we need to find a square of 
 the polynomials:
\be
Q_n^2 = {\det}_n \left( C_{ij} \right),
\ee
where
\bel{CMM}
C_{jk} = ( M^T M )_{jk} = 
\sum_{i=1}^n \s_{3i-2j-1} \s_{3i-2k-1} = 
\sum_{i=-\infty}^n \s_{3i-2j-1} \s_{3i-2k-1}.
\ee
Here we used the relation $\sigma_k = 0$ for $k<0$.

Note that the elementary symmetric polynomials have the following expression:
\be
\s_k = \frac{1}{2 \pi i} \oint_{\gamma} \frac{dz}{z^{n - k + 1}}
\prod_{m = 1}^n ( z + x_m ),
\ee
where an integration contour $\gamma$ is a circle around origin in
positive direction.

Substituting the above expression into \Eq{CMM} and summing up
an infinite series, we have
\be
C_{jk} = \oint_{\gamma_1} \frac{dz_1}{2\pi i}
\oint_{\gamma_2} \frac{dz_2}{2\pi i}
\frac{ z_1^{2n-2j+1} z_2^{2n-2k+1}}{ (z_1 z_2)^3 - 1 }
\prod_{m = 1}^n ( z_1 + x_m )( z_2 + x_m ).
\ee
Here we have chosen the radius of the circle $\gamma_i$ ($i = 1, 2$)
to be  greater than one in order for the series to converge.

The matrix $C$ still depends on $n^2$ different functions $C_{jk}$.
However, this matrix can be transformed to a more convenient form.
Let us introduce the following matrix
\be
A_{jk}=\frac{1}{(n - j)!}
\left. \frac{d^{n - j}}{d(x^2)^{n - j}}
\prod\limits_{m \ne k}^n \left( x^2 + x_m^2 \right) \right|_{x^2 = 0},
\ee
with a determinant
\be
\det A=\prod_{i<j}^n( x_i^2 - x_j^2 ).
\ee
Using this matrix, we define another matrix $D$,
which differs from $C$ by linear transformation
\be
D = A^T C A.
\ee
We have an explicit expression for matrix elements of D:
\be
D_{jk} = \frac{1}{(2\pi i)^2}
\oint \ d^2z \frac{z_1 z_2}{z_1^3 z_2^3-1}
Y(z_1,x_j)Y(z_2,x_k),
\ee
where
\be
Y(z, x)=\frac{J(z)}{z^2 + x^2},
\ee
and
\be
J(z) = \prod_{a = 1}^n ( z + x_a ) ( z^2 + x_a^2 ).
\ee
Taking the integral with respect to $z_2$,
we have after the symmetrization of the integrand
\bel{Djk}
D_{jk}=\frac1{18\pi i}\oint_{\gamma}\frac{dz}{z}
\Bigl(\sum_{l=1}^3 \omega^l Y(\omega^{-l} z, x_j)\Bigr)
\Bigl(\sum_{m=1}^3 \omega^m Y(\omega^{-m} z^{-1}, x_k)\Bigr).
\ee
Here $\omega = e^{2\pi i/3}$ and the integration contour $\gamma$ is a 
circle whose radius is larger than one. 
Explicitly, the sum in $l$ can be written as
\be
\sum_{l=1}^3 \omega^l Y(\omega^{-l} z, x) = 
Y(z, x) + \omega Y(\omega^{-1} z, x) + \omega^{-1} Y(\omega z, x).
\ee

Determinants of matrix $C$ and $D$ are related by
\be
{\det}_n C = \prod_{i < j}^n ( x_i^2 - x_j^2)^{-2}{\det}_n D.
\ee
Thus, we obtain a determinant representation for $Q_n^2$:
\be
Q_n^2 = \frac{{\det}_nD}
{\prod\limits_{i<j}^n ( x_i^2 - x_j^2 )^2}.
\ee

\resection{Dual fields}

Entries of the matrix $D_{jk}$ are parameterized now by a single function
$D$. However, an element $D_{jk}$,  is still  not a
 function of two arguments only, because of  the product
$J(z)=\prod_{m=1}^n( z + x_m )( z^2 + x_m^2 )$. 
This product depends on all  $x_m$. In order to get rid of these products we
introduce auxiliary Fock space and auxiliary  quantum operators---dual fields.

Let us define
\bel{Ddef}
\begin{array}{l}
\Phi_1(x)=q_1(x)+p_2(x), \qquad
\Phi_2(x)=q_2(x)+p_1(x),
\end{array}
\ee
where operators $p_j(x)$ and $q_j(x)$ act in the canonical Bose Fock space 
in a following way
\be
(0| q_j(x)=0, \qquad p_j(x)|0) = 0.
\ee
Non-zero commutation relations are given by
\bel{Dcom}
[p_1(x),q_1(y)]=[p_2(x),q_2(y)]=\xi(x,y)=
\log\Bigl(( x + y )( x^2 + y^2 )\Bigr).
\ee
Due to the symmetry of the function $\xi(x,y)=\xi(y,x) $,
all fields $\Phi_j(x)$ commute with each other
\be
[\Phi_j(x),\Phi_k(y)]=0,\qquad j,k=1,2.
\ee
Dual fields are linear combinations of canonical Bose fields , see page 210
of \cite{K.B.I.}.

Instead of $Y(z,x)$ let us define an operator valued function
\be
\hat{Y}(z,x)=\frac{e^{\Phi_1(z)}}{ z^2 + x^2 } .
\ee
Instead of $D_{jk}$ \Eq{Djk}, we shall introduce an operator in the
auxiliary Fock space: 
\bel{hatD}
\hat{D}(x, y)=\frac1{18\pi i}\oint\frac{dz}{z}
\Bigl(\sum_{l=1}^3 \omega^l \hat{Y}(\omega^{-l} z, x)\Bigr)
\Bigl(\sum_{m=1}^3 \omega^m \hat{Y}(\omega^{-m} z^{-1}, y)\Bigr).
\ee
It is easy to show that an exponent of dual field acts like a shift
operator. Namely, if $g\left({\Phi_1(y)}\right)$ is a
function of $\Phi_1(y)$ then
\[
(0| \prod_{m = 1}^n e^{\Phi_2(x_m)} g\left({\Phi_1(y)}\right) |0) = 
(0| g\left({q_1(y)+\sum_{m = 1}^n \xi(x_m, y)}\right) |0)
= g(\log J(y)).
\]
Using this property of dual fields one can remove the products
$J(z)$ from the matrix $D_{jk}$.
For more detailed derivation one should look in at formula (3.6) of the
paper \cite{KS}.

Standard arguments of quantum field theory show that 
\be
{\det}_nD = (0|
{\det}_n\left(\hat{D}(x_j,x_k)e^{\frac12\Phi_2(x_j)+\frac12\Phi_2(x_k)}
\right)|0).
\ee

Up to now, we wrote $Q_n^2$ factor of $|F_n|^2$ as a determinant.
An absolute value of the form factor is equal to 
\ba
& &|F_n(\beta_1,\dots,\beta_n)|^2 \nn
&=&
|H_n|^2 Q_n^2 \prod_{i<j} \left|
\frac{ f(\beta_i - \beta_j)}{( x_i + x_j )} \right|^2 \nn
&=&
|H_n|^2 {\det}_n D \prod_{i<j}
\left|
\frac{f(\beta_i - \beta_j)}{( x_i^2 - x_j^2 )( x_i + x_j )}
\right|^2.
\ea
In order to factorize the double product part, we introduce another
dual field
\bel{Ddef0}
\tilde\Phi_0(x) = \tilde q_0(x) + \tilde p_0(x).
\ee
As usual
\be
(0| \tilde q_0(x)=0, \qquad \tilde p_0(x)|0) = 0.
\ee
Operators $\tilde q_0(x)$ and $\tilde p_0(y)$  commute with all  $p_j$ and 
$q_j$ ($j=1,2$). The only non-zero  commutation relation is 
\bel{Dcom0}
[\tilde p_0(x),\tilde q_0(y)] = \eta(x,y),
\ee
where
\bel{eta0}
\eta(x,y)=\eta(y,x)
=2\log\left|\frac{f(\log\frac xy)}{( x^2 - y^2)(x + y)}\right|.
\ee
Here we have used the fact that $|f(\beta)|$ is a symmetric function : 
$|f(-\beta)|=|f(\beta)|$ .
It is worth mentioning also that the right hand side of \Eq{eta0} has no
singularity at $x=y$, because $f(\beta)$ has the first order zero at
$\beta = 0$.
Using  some of equations for minimal form factor from  \cite{Z}
\be
f(\beta)f(\beta + i \pi) = \frac{\sinh \beta}{\sinh \beta - i\sin
(\pi/3)},
\ee
and $f(i \pi) = 4 v^2(0)$, we can see that
\be
f'(0)=\frac{i}{2\sqrt{3} v^2(0)}.
\ee
Hence
\bel{Detaxx}
\eta(x,x)=-2\log|\lambda x^3|,
\ee
where
\be
\lambda = 8\sqrt{3} v^2(0).
\ee
Newly introduced dual field also commute
\bel{DAbel0}
[\tilde\Phi_0(x),\tilde\Phi_0(y)] = 0 =
[\tilde\Phi_0(x),\Phi_j(y)].
\ee
Due to the Campbell--Hausdorff formula, we have
\bel{Dvac0}
(0| \prod_{m=1}^ne^{\tilde\Phi_0(x_m)} |0) =
\prod_{i,j=1}^n e^{\frac12\eta(x_i,x_j)}
=\lambda^{-n}
\prod_{m = 1}^n x_m^{-3}\prod_{i<j}^n
\left|\frac{f(\log\frac{x_i}{x_j})}{(x_i^2 - x_j^2)(x_i + x_j)}\right|^2.
\ee
Combining these results, we can  represent  a square of an
absolute value of the form factor as a determinant:
\ba
& &|F_n(\beta_1, \ldots, \beta_n)|^2 \nn
&=&
\left(\frac{\pi m^2}{4\sqrt{3}}\right)^2 12^n
(0|{\det}_n\left( x_j^{3/2} x_k^{3/2} \hat{D}(x_j, x_k) 
e^{\frac{1}{2}\Phi_0(x_j) + \frac{1}{2}\Phi_0(x_k)} 
\right) |0). \nonumber
\ea
Here
\be
\Phi_0(x) = \tilde\Phi_0(x) + \Phi_2(x).
\ee
So we managed to represent a square of an absolute value of the
form factor as a determinant, similar to one of the terms in the 
right hand side of \Eq{TdefFrdet}. 
In the next section we shall sum up all contributions of the  form
factors and obtain a determinant representation for the correlation function. 

\resection{The determinant representation}

Because the scaling Lee-Yang model is a non-unitary theory, 
the normalization constant $H_n$ \Eq{Hn} is pure imaginary for $n$
odd. This leads to the following relation:
\be
\langle \vac| \Theta(0) | \beta_1, \ldots, \beta_n \rangle
\langle \beta_n, \ldots, \beta_1 | \Theta(0) |\vac \rangle
=(-1)^n | F_n (\beta_1, \ldots, \beta_n)|^2.
\ee
Then the correlation function of $\Theta$ can be written as
\ba
& & \langle \Theta(x) \Theta(0) \rangle \nn
&=&
\sum_{n=0}^\infty (-1)^n \int\frac{d^n\beta}{n!(2\pi)^n} 
|F_n(\beta_1,\dots,\beta_n)|^2\prod_{j=1}^ne^{-\theta(x_j)} \nn
&=&
\left(\frac{\pi m^2}{4\sqrt{3}}\right)^2
(0| \sum_{n=0}^{\infty} \int_0^{\infty} 
\frac{d^n x}{n!}\left(-\frac{6}{\pi}\right)^n \nn
& & \qquad \times {\det}_n
\left(x_j x_k \hat{D}(x_j, x_k)
e^{\frac{1}{2}(\Phi_0(x_j) + \Phi_0(x_k))} 
e^{-\frac{1}{2}( \theta(x_j) + \theta(x_k) )}
\right) |0). \nonumber
\ea
where
\be
\theta(x)=\frac{mr}2(x+x^{-1}).
\ee
Now we can use \Eq{TdefFrdet} in order to sum up.
Finally  we obtained  a determinant representation in terms of an integral
 operator (Fredholm determinant)
\bel{Ldet}
\langle \Theta(x) \Theta(0) \rangle = 
\left(\frac{\pi m^2}{4\sqrt{3}}\right)^2
(0| {\det}( I - \hat{U} ) |0).
\ee
The operator $\hat{U}$ act on a function $g(x)$ as
\be
[\hat{U}g](x) = \int_0^{\infty} dy \ \hat{U}(x, y) g(y).
\ee
The kernel of the integral operator $\hat{U}(x,y)$ is equal to
\bel{Lkern}
\hat{U}(x, y) = \frac{6}{\pi} x y
\hat{D}(x, y) 
e^{\frac{1}{2}(\Phi_0(x) + \Phi_0(y))} 
e^{-\frac{1}{2}( \theta(x) + \theta(y) )},
\ee
where $\hat{D}$ is given by \Eq{hatD}
\be
\hat{D}(x, y)=\frac1{18\pi i}\oint\frac{dz}{z}
\Bigl(\sum_{l=1}^3 \omega^l \hat{Y}(\omega^{-l} z, x)\Bigr)
\Bigl(\sum_{m=1}^3 \omega^m \hat{Y}(\omega^{-m} z^{-1}, y)\Bigr),
\ee
where
\be
\hat{Y}(z,x)=\frac{e^{\Phi_1(z)}}{ z^2 + x^2 }.
\ee
The dual fields $\Phi_0(x)$ and $\Phi_1(x)$ were defined in the
section 3 (see \Eq{Ddef} and \Eq{Ddef0}).
The main property of these dual fields is,
that they commute with each other, so the Fredholm determinant
$\det(I - \hat U)$ is well defined.
Certainly $\det(I - \hat{U})$ is an operator in auxiliary Fock space,
but it belongs to the Abelian sub algebra.
On the other hand, the vacuum expectation value of these operators is
non-trivial. It follows from commutation relations \Eq{Dcom},
\Eq{Dcom0}, that in order to calculate the vacuum expectation value,
one should use the following prescription
\be
(0| \prod_{a=1}^{M_1}e^{\Phi_0(x_a)}
\prod_{b=1}^{M_2}e^{\Phi_1(x_b)} |0)=
\prod_{a=1}^{M_1}\prod_{b=1}^{M_1}e^{\frac12\eta(x_a,x_b)}
\prod_{a=1}^{M_1}\prod_{b=1}^{M_2}e^{\xi(x_a,x_b)}.
\ee
Here
\be
\eta(x,y)=2\log\left|\frac{f(\log\frac{x}{y})}{(x^2 - y^2)(x+y)}\right|,
\ee
and
\be
\xi(x,y)=\log\Bigl((x + y)(x^2 +y^2)\Bigr).
\ee
Similar Fredholm determinant representations were useful in the
theory of correlation functions \cite{K.B.I.,EFIK,ES,KS1}.

\resection{Large $r$-asymptotic}

Using the technique of \cite{KS}, we calculate the long distance
asymptotic of the correlation function.

The determinant can be written as
\be
{\det}(I-\hat{U}(x,y)) = {\det}(I-\tilde{U}(z_1, z_2)),
\ee
where
\bel{tilU}
\tilde{U}(z_1, z_2) = \int_0^{\infty} dx P_1(z_1,x) P_2(z_2, x),
\ee
and
\be
P_1(z, x) = \frac{x}{3\pi^2 i z}\left(\sum_{l = 1}^3 \omega^l
\hat{Y}(\omega^{-l} z, x)\right) 
e^{\frac{1}{2}\Phi_0(x) - \frac{1}{2}\theta(x)},
\ee
\be
P_2(z, y) = y\left(\sum_{l = 1}^3 \omega^l
\hat{Y}(\omega^{-l} z^{-1}, y)\right) 
e^{\frac{1}{2}\Phi_0(y) - \frac{1}{2}\theta(y)}.
\ee
The integral operator $\tilde{U}(z_1, z_2)$ acts on a function
$g(z)$ as
\be
[\tilde{U}g](z_1) = \oint \tilde{U}(z_1, z_2) g(z_2) dz_2.
\ee
Here the integration contour is a circle around zero in positive
direction.

In the limit $r \rightarrow \infty$, we evaluate the integral
\Eq{tilU} by a saddle point method. The saddle point of the function
$\theta(x)$ is $x=1$.
Hence, we can estimate the integral in \Eq{tilU} as
\bel{LtUest}
\tilde U(z_1, z_2)=
P_1(z_1, 1)P_2(z_2, 1)\left(\sqrt{\frac{2\pi}{mr}}+O(r^{-3/2})\right).
\ee
Thus, for the large $r$
asymptotics the kernel $\tilde U(z_1,z_2)$ becomes
a one-dimensional projector, and its Fredholm determinant is 
equal to 
\bel{Lasydet} 
\det(I - \tilde U)
\to 1 - \oint dz \ \tilde U(z,z).  
\ee 
Using the commutation relations of dual fields, 
we can evaluate $(0| \tilde{U}(z, z) |0)$ as follows:
\be
(0| \tilde{U}(z, z) |0)=\frac{e^{-mr}}{3\pi^2i \lambda z}
\sqrt{\frac{2\pi}{mr}} Y_1(z) Y_1(z^{-1}) + \ldots,
\ee
where
\be
Y_1(z) = \sum_{l=1}^3 \omega^l (\omega^{-l} z + 1) = 3z.
\ee 
The above result leads to the asymptotic form of the vacuum mean value 
of the Fredholm determinant
\be
(0| {\det}(I-\tilde{U}) |0) = 1 - \frac{\sqrt{3}}{2v^2(0)}
\frac{e^{-mr}}{\sqrt{2\pi m r}} + \ldots ,
\ee
which agrees with the large distance contribution from
zero- and one-particle states \cite{Z}.

\section*{Summary}

We considered scaling Lee-Yang Model and we obtained determinant 
representation for the correlation function of the trace of the 
energy-momentum tensor. We believe that determinant representations is 
the universal language for the description of quantum correlation 
functions in the integrable (exactly solvable) models of quantum field
theory. 

\section*{Acknowledgments}

We are grateful to A. Berkovich and N. Slavnov for useful discussions.
This work was supported  by NSF Grant No PHY-9605226.

\end{document}